\documentclass[11pt]{article}
\usepackage{graphicx}
\begin{document}

\title{Nanoelectromechanical systems based on multi-walled
nanotubes: nanothermometer, nanorelay and nanoactuator}

\maketitle

\begin{center}
Andrey M.Popov\footnote{e-mail: popov@ttk.ru},  Yurii E.Lozovik

 Institute of Spectroscopy,
Troitsk, Moscow region, 142190, Russia

Elena Bichoutskaia\footnote{e-mail: Elena.Bichoutskaia@nottingham.ac.uk}

Department of Chemistry, University of Nottingham, University Park, Nottingham, NG7 2RD, UK

Anton S.Kulish

Moscow Institute of Steel and Alloys, Leninskii prosp. 4, Moscow,
117936, Russia
\end{center}

\begin{abstract}
We report on three new types of nanoelectromechanical systems
based on carbon nanotubes: an electromechanical nanothermometer, a
nanorelay and a nanomotor. The nanothermometer can be used for
accurate temperature measurements in spatially localized regions
with dimensions of several hundred nanometers. The nanorelay is a
prototype of a memory cell, and the nanoactuator can be used for
transformation of the forward force into the relative rotation of
the walls. Relative motion of the walls in these nanosystems is
defined by the shape of the interwall interaction energy surface.
{\it Ab initio} and semi-empirical calculations have been used to
estimate the operational characteristics and dimensions of these
nanosystems.
\end{abstract}

\section{Introduction}
Unique properties of carbon nanotubes, such as free sliding and rotation of the walls and metallic conductivity, allow using the walls of
nanotubes as movable elements and elements of electric circuit in nanoelectromechanical systems (NEMS). An example of such applications is a
nanotube-based nanomotor which has been produced recently \cite{fennimore03,bourlon04}. A set of NEMS based on relative motion of the walls of
nanotubes, such as a nanospring \cite{cumings00}, a nanoswitch \cite{forro00} and a gigahertz oscillator \cite{zheng02}  have been suggested in
the literature. A number of nanorelays with one electrode made of a carbon nanotube have also been considered
\cite{dequesnes02,kinaret03,maslov06}. A new type of NEMS in which the relative motion of the walls is controlled by the shape of the interwall
interaction energy surface has been proposed in \cite{lozovik03,lozovik03a} as a nanodrill, in \cite{bichoutskaia06a} as a tension nanosensor
and in \cite{belikov04,bichoutskaia06} as rotational nanobearings with fixed position along the axis. In this paper, we propose three new NEMS
of this type: an electromechanical nanothermometer, a nanorelay with all electrodes made of nanotubes, which can also be used as a memory cell,
and a nanoactuator for transformation of forward force into relative rotation of the walls.

\section{Methods of calculations}
The density functional theory supercell code within the local density approximation (AIMPRO
\cite{briddon00}) has been used to calculate the interwall interaction energy and the barriers to
the relative sliding of the walls for the (5,5)@(10,10) and (6,6)@(11,11) double-walled carbon nanotubes (DWNTs). The details of the method can be found in \cite{bichoutskaia06,bichoutskaia05}. Within AIMPRO, the pseudowave functions are described by 4 atom-centered Gaussian functions per atom expanded in spherical harmonics up to $l$=1, with the second smallest exponent expanded to $l$=2. The supercell consists of 60 and 68 carbon atoms for the (5,5)@(10,10) and (6,6)@(11,11) DWNTs, respectively. The Brillouin Zone sampling has been performed using 18 special $k$-points in the direction of the nanotube axis. Non-local, norm-conserving pseudopotential and the standard
Perdew-Wang exchange-correlation functional \cite{perdew} have been used. Minima in the total energy are found using a conjugate gradient scheme to an accuracy of 1 $\mu$eV/atom. Positions of all atoms in the isolated (5,5), (6,6), (10,10) and (11,11) walls have been optimized.

Structures of the caps of the nanotubes used as the electrodes of a nanorelay have been calculated using the Q-Chem 2.1 quantum chemistry package \cite{q-chem}. The 6-12 Lennard-Jones potential
$U$=$4\epsilon((\sigma/r)^{12}-(\sigma/r)^{6})$ has been used for calculation of the interaction between the electrodes of a nanorelay. Parameters  $\sigma$=3.44 \AA and $\epsilon$=$2.62\cdot10^{-3}$ eV \cite{girifalco00}) have been used for the interaction between carbon atoms, and $\sigma$=2.039 \AA,
 $\epsilon$=0.14438 eV \cite{ellis00}  for the interaction between carbon and copper atoms.

\section{Nanoelectromechanical systems}

\subsection{Electromechanical nanothermometer}

A new concept of an electromechanical nanothermometer is based on
the interaction and relative motion of the walls of carbon
nanotubes. Temperature measurements are carried out through the
measurements of the electrical conductance of the walls as a
function of their relative displacement. Relative position of the
walls of a nanotube depends on and changes with the temperature
due to the thermal vibrations of the walls. The probability of the
relative displacement of the walls, $p(q)$, is defined by the
energy of their interaction, $U(q)$, as $p(q) \sim exp(-U(q)/kT)$.
For a given temperature, the total conductance, $G_{tot}(T)$, can,
therefore, be found as the expectation value of the conductance at
a fixed position of the walls, $G(q,T)$ \cite{degroot}

\begin{equation}\label{gtot}
G_{tot}(T) = \frac {\int \limits^{\infty}_{-\infty}
G(q,T)\exp(-U(q)/kT)dq} {\int \limits^{\infty}_{-\infty}
\exp(-U(q)/kT)dq}.
\end{equation}

\noindent Equation \ref{gtot} takes into account contributions
from the thermal vibrations of the walls.

The working conditions for the electromechanical nanothermometer are the following: $G(q,T)$ depends significantly on the coordinates $q$ (condition $A$); $G(q,T)$ depends weakly on the temperature $T$ (condition $B$) (contributions from phenomena other than the thermal vibrations of the walls are insignificant); amplitude of the thermal vibrations is large enough to provide the main contribution to the temperature dependence of the total conductivity $G_{tot}(T)$ (condition $C$); amplitude of the thermal vibrations is still small for the relative displacements of the walls to upset the normal operation of the system (condition $D$).

A model implementation of the nanothermometer is based on the
(6,6)@(11,11) DWNT (Figure 1). The conductance of the
(6,6)@(11,11) DWNT depends significantly on the relative position
of the walls in both telescopic and shuttle systems \cite{grace04}
and changes due to the thermal vibrations of the walls. Therefore,
the condition $A$ is fulfilled. Experimental studies of
\cite{gao05} show that the conductance of single-walled carbon
nanotubes depends weakly on the temperature in the region of
$T>$50 K. Theoretical studies of \cite{lebedev06} conclude that
for DWNTs with fixed walls it depends weakly on the temperature
for $T>$100 K. Thus, the condition $B$ is also satisfied in these
temperature regions.

\begin{figure}[htb]
\includegraphics[width=\textwidth]{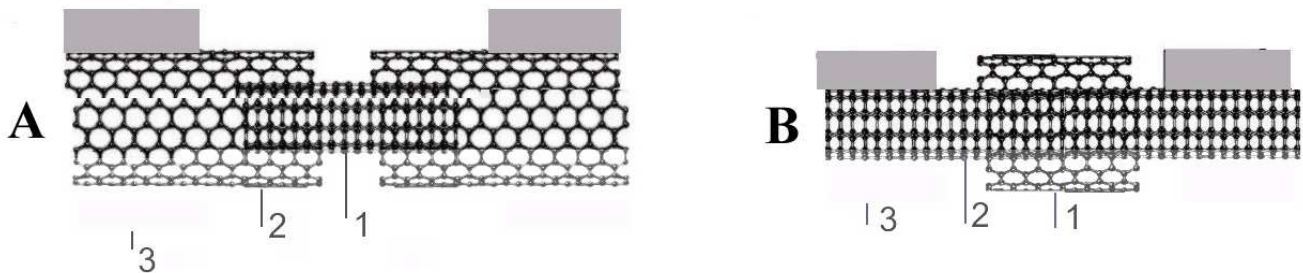}
\caption{Schematic of an electromechanical nanothermometer based on DWNT. A: a shuttle nanothermometer with the movable inner wall, B: a telescopic nanothermometer with the movable outer wall. 1 is the movable wall, 2 is the fixed wall with the attached electrodes 3. } \label{fig:1}
\end{figure}

The dependence of both the interaction energy of the walls of a
nanotube and the conductance on the displacement of the walls is
uniquely determined by the symmetry of  the (6,6)@(11,11) DWNT.
$U(q)$ and $G(q)$ show some common features, for example, they
have the same period of oscillation and the extrema in both
functions coincide. Moreover, the dependence of the interwall
interaction energy and the conductance on the angle of relative
rotation of the walls is negligible due to incompatibility of the
rotational symmetries of the (6,6) and (11,11) walls.

In the absence of diffusion of the short wall, integration in
equation (\ref{gtot}) can be performed near the bottom of the
potential well of the interwall interaction energy $U(q)$. This
region can be interpolated as \cite{therm}

\begin{equation}\label{expansion4}
               U(q')=U_1+\frac{\pi \Delta U_q}{\delta_q^2} q'^2,
\end{equation}

\noindent where $U_1$ is the interwall interaction energy of the
ground state, $\Delta U_q$ is the energy barrier between the
minima of $U(q)$, $\delta_q$ is the period of displacement of the
walls between the minima,  and $q'$ is a displacement of the
movable wall from the position which defines the ground state. For
small values of $q'$, $G(q',T)$ can be interpolated as
\cite{therm}

\begin{equation}\label{gz}
               G(q')=G_1(T)(1+\gamma q'^2),
\end{equation}

\noindent where $G_1$ is the conductivity of DWNT in the ground
state. The value of $\gamma$ for the overlap of the walls of 2.45
nm can be extracted from the Figure 3 of \cite{grace04} as $\gamma
\sim $850 \AA$^{-2}$.

Substitution of (\ref{expansion4}) and (\ref{gz}) into
(\ref{gtot}) leads to the following expression for the dependence
of the conductance of the nanothermometer on the temperature

\begin{equation}\label{gtot2}
G(T) = G_1(T)(1+\frac{\gamma \delta_q^2 kT}{\pi \Delta U_q})=G_1(T)(1+HT).
\end{equation}

Here the energy barrier $\Delta U_q$ for the (6,6)@(11,11) DWNT
has been calculated {\it ab initio} and for the overlap of the
walls of 2.45 nm it has the value of 1.96 eV. Therefore, for T=300
K, the value of $HT$ in (\ref{gtot}) is about 45 and this ensures
the fulfillment of the condition $C$. For the condition $D$ to be
satisfied, the displacement of the shuttle as a result of
diffusion has to be less than the distance between the electrode
and the shuttle. The minimum length between the electrodes of the
shuttle nanothermometer which operates during 100 years without a
failure is estimated as 38 nm using the {\it ab initio} results
\cite{bichoutskaia06b} for the diffusion coefficient of the walls
of (6,6)@(11,11)DWNT. The nanothermometer can be calibrated with
the use of a thermocouple, and since the temperature measurement
is based on conductance measurement, the same order of accuracy as
in the case of thermocouple using can be achieved (for example,
the accuracy of measurements at room temperature is about 0.1 K
for copper-constantan and chromel-alumel thermocouples).

\subsection{Nanorelay}

The essential working condition for a nanosystem to operate as a
relay is the presence of two minima in the interaction energy of
its components (positions 'on' and 'off'), or so called
bi-stability. The new type of nanotube-based nanorelay has been
proposed recently (Figure 2A). Here we proposed new schemes of
nanorelays of this type (Figures 2B and 2C). Their operation is
fully determined by the the forces applied to the movable inner
wall 1. These forces are the attraction force, $F_a$, of the Van
der Waals interaction between the electrodes; the electrostatic
force, $F_e$, between the electrodes 3 and 4 (Figures 2A and 2B)
or between the electrode 3 and the control electrode 5 (Figure
2C); and the forces of the interwall interaction, such as the
capillary force $F_c$, which retracts the inner wall, and the
static friction force $F_f$.

For the (5,5)@(10,10) DWNT, the capillary force $F_c$ is estimated
in \cite{bichoutskaia06} as 0.625 nN. The results for calculations
of the interaction between the electrodes are presented in Table
1. According to the {\it ab initio} \cite{bichoutskaia05} and
semiempirical \cite{belikov04} calculations, the friction force
$F_f$ is approximately zero for DWNTs with incommensurate or/and
chiral walls. If the condition for bi-stability, $F_a > F_c$, is
satisfied then the nanorelay can be used as a memory cell of the
external storage. If one of the electrodes  of the nanorelay is
made of a DWNT with incommensurate or chiral walls, the condition
for bi-stability can only be achieved if the second electrode is
flat (see Figure 2A). An 'all-carbon' nanorelay with both
electrodes made of carbon nanotubes is shown in Figure 2B. In the
case of small friction force $F_f$, this nanorelay is stable in
position 'on' only if the voltage is applied. Therefore, it can be
used as an easy-to-operate memory cell of the on-line storage (see
Figure 2D).

\begin{figure}[htb]
\includegraphics[width=\textwidth]{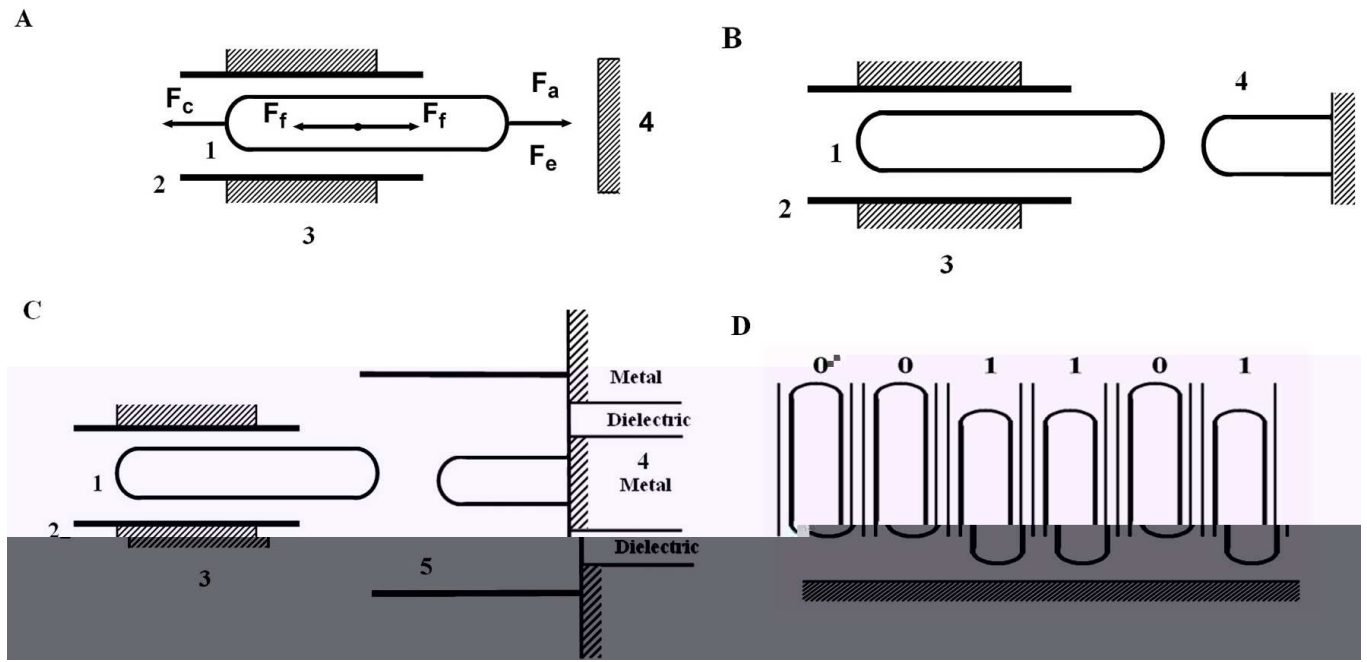}
\caption{Schematics of memory cells based on DWNTs (in position
'on'). A: a memory cell with a flat second electrode; B: an
'all-carbon' memory cell with a carbon nanotube attached to the
second electrode; C: an 'all-carbon' memory cell with a control
electrode made of a nanotube. 1 is the movable inner wall; 2 is
the fixed outer wall; 3, 4 and 5 are the first, second and the
control electrodes. Figure D shows an example of the storage
produced from memory cells.} \label{fig:2}
\end{figure}

In the case of DWNTs with nonchiral commensurate walls, the
friction force $F_f$ is significant
\cite{belikov04,bichoutskaia05}. For the (5,5)@(10,10) DWNT, the
dependence of the force $F_f$ on the overlap of the walls,
$l_{ov}$, can be described as $F_f=Al_{ov}\cos(2\pi
l_{ov}/\delta_{q})$, where $A$ = 0.01 N/m \cite{bichoutskaia06}.
The analysis of the balance of forces in 'all-carbon' nanorelay
shows that for the critical overlap $l_c$ = 4 nm, the sum of the
interwall static friction force and the attraction force of the
Van der Waals interaction between the electrodes, $F_f + F_a$, has
to be greater than the capillary force $F_c$. If the overlap of
the walls is greater than the critical value, $l_{ov} > l_c$, the
nanorelay can be bi-stable.

The condition $l_{ov} > l_c$ is not sufficient for the successful
operation of the 'all-carbon' nanorelay as a memory cell of the
external storage. The energy barrier between positions 'on' and
'off' can be, in principle, overcome as the result of the thermal
diffusion of the inner wall 1. The probability of this to happen
is determined by the Arrhenius formula $p=\Omega \exp(-\Delta
U/kT)$, where $\Omega$ is the frequency multiplier. According to
the molecular dynamics simulation of the reorientation of shells
of the C$_{60}$@C$_{240}$ nanoparticle \cite{lozovik00}, the
frequency multiplier is about an order of magnitude greater than
the frequency of small oscillations near the minimum of the
intershell interaction energy. The dependence of the lifetime,
$t=p^{-1}$, of the 'all-carbon' nanorelay in position 'on'  on the
overlap of the walls is shown on in Figure 3.

For the case of a nanorelay with the control electrode made of a
carbon nanotube (Figure 2C), the electrostatic force $F_e$ acting
on the (5,5) movable wall has been calculated. The diameter of the
movable wall, as well as the control electrode, is taken to be 3
nm as the armatures of cylindrical capacitor. Assuming the
friction force $F_f$ is negligible, the voltage of the switching
between positions 'off' and 'on' can be determined by the
condition $F_e > F_c$ and estimated to be about 6 V. In this case,
the voltage needed to hold the nanorelay in position 'on' is about
4.8 V.

\begin{figure}[t]
\includegraphics[width=\textwidth]{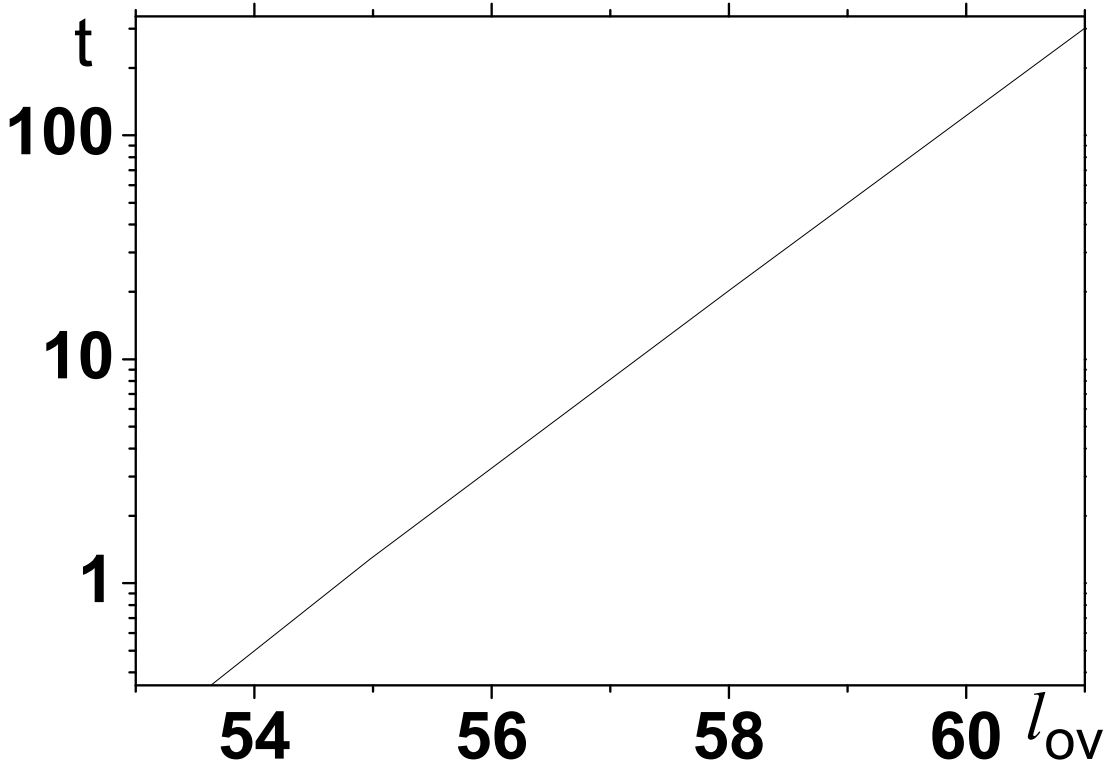}
\caption{The dependence of the lifetime of the 'all-carbon'
nanorelay in position 'on'  on the overlap of the walls. The
nanorelay contains the (5,5)@(10,10) DWNT as the first electrode
and the (5,5) nanotube as the second electrode.} \label{fig:3}
\end{figure}

\begin{table}
\caption{Characteristics of the interaction between the
electrodes of the nanorelay: $U_{min}$ (in eV) is the minimum
value of the interaction energy between the electrodes, $R_{min}$
(in \AA ) is the distance between the electrodes which corresponds
to the minimum of the interaction energy; $F_a^{max}$ (in nN) is
the maximum attraction force between the electrodes.}
\label{tab:4}\renewcommand{\arraystretch}{1.5}
\begin{tabular}{lccc} \hline
second electrode & $U_{min}$ & $R_{min}$ & $F^{max}_a$ \\ \hline
graphite & -0.914 & 3.153 & 0.683 \\
copper & -35.69 & 1.879 & 42.32\\
(5,5) nanotube & -0.259 & 3.345 & 0.233\\ \hline
\end{tabular}
\end{table}

\subsection{Nanoactuator}

A nanoactuator has been designed for transformation of the forward
force into the relative rotation of the walls. Schematic of the
nanoactuator is shown in Figure 4. The wall 1 of the nanotube is
fixed making a stator. Position of a rotor (the walls 2 and 3) is
also fixed. The walls 3 and 4 make up a bolt-nut pair which
converts the motion driven by an axially directed force applied to
the wall 4 into the rotation of the rotor. It has been shown in
\cite{lozovik05,lozovik06} that the bolt-nut pair can be a DWNT
with periodically positioned defects.

The first of the essential working conditions for the nanoactuator
is defined by the motion of the walls 1 and 2, which constitute a
rotational nanobearing. This condition states that the barrier to
the rotation of the walls 1 and 2 has to be low and the barrier to
the sliding axial motion of these walls has to be high. This
condition is satisfied if both 1 and 2 are commensurate achiral
walls with incompatible rotational symmetry
\cite{belikov04,bichoutskaia06}.

The second of the essential working conditions for the
nanoactuator is defined by the motion of the walls 3 and 4, which
constitute a bolt-nut pair. If a short impulse of the force
directed along the axis is applied to the wall 4, this wall
acquires the kinetic energy sufficient to overcome the barrier
$E_1$ to the motion of the walls 3 and 4 along the line of the
thread. However, the energy will not be large enough to overcome
the barrier $E_2$ to the motion of the walls 3 and 4 across the
line of the thread, i.e.

\begin{equation}
\label{neq}
          \frac{M {\bf V}^2\sin^2 \chi}{2} > E_1,  ~~~~~~~~~~~~~~~~~~~~~~ \frac{M {\bf V}^2\cos^2 \chi}{2} < E_2,
\end{equation}

\noindent where $M$ and ${\bf V}$ are the mass and velocity of the
wall 4, and $\chi$ is the angle of the thread. The inequalities
(\ref{neq}) are fulfilled if $\cot^2 \chi<E_2/E_1=\beta$ where
$\beta$ is the depth of the thread as given in
\cite{lozovik03,lozovik03a}. Evidently, if the angle $\chi$ of the
thread is greater than 45 degrees than the nanoactuator can
operate with any depth of the thread. The charges on the edges of
the wall 4 can be produced by chemical adsorption.

The conservation of the angular momentum of the nanoactuator gives the minimum length $L^{min}_3$ of the
wall 3 for which the rotation at an angle $\xi$ becomes possible

\begin{equation}
\label{l3}
    L^{min}_3=\frac{L_4^2 R_4^2}{L_4 R_4^2 + \xi (R_2^3 + R_3^3) \sin \chi},
\end{equation}

\noindent where $L_4$ is the length of the wall 4; $R_2$, $R_3$
and $R_4$ are the radii of the walls 2, 3 and 4, respectively. The
operation time of the proposed nanorelay and nanoactuator can be
the same order of magnitude as oscillation period $10-100$ ps of
nanotube-based gigaherts oscillator \cite{zheng02}.

\begin{figure}[htb]
\includegraphics[width=\textwidth]{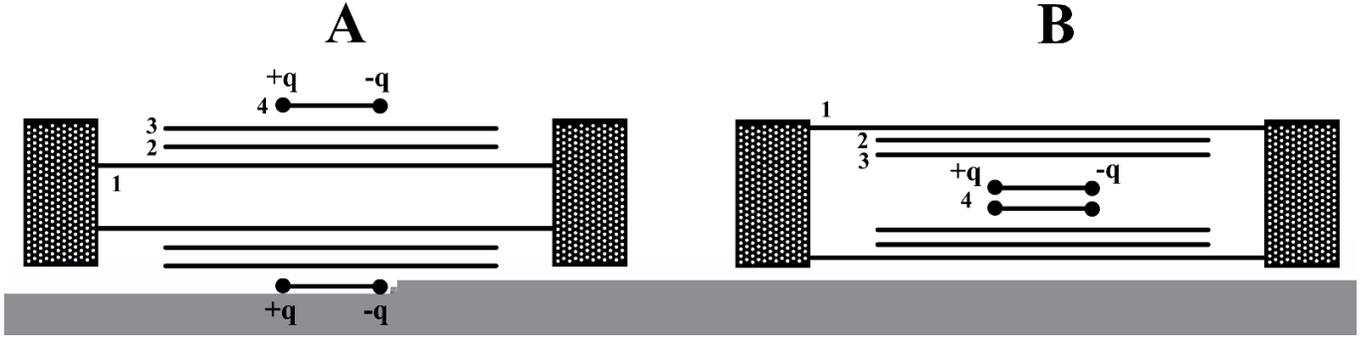}
\caption{Schematic of a nanoactuator. A: a nanoactuator with the
inner wall 1 as a stator, B: a nanoactuator with the outer wall 1
as a stator. The walls 1 and 2 comprise the rotational
nanobearing, the walls 3 and 4 are the bolt-nut pair with the
thread-like shape of the interwall interaction energy.}
\label{fig:4}
\end{figure}

\section{Conclusive remarks}

We proposed three new nanotube-based NEMS: a nanothermometer, a
nanorelay with 'all-carbon' electrodes and a nanoactuator. A
considerable progress has been recently achieved in the
nanotechnology techniques in the field of production of NEMS. A
nanomanipulator can be now attached to MWNT in order to move
individual  walls \cite{cumings00}, manipulation with nanotubes is
routinely possible \cite{yu99}, the caps of the walls can be
removed \cite{tsang93}, nanotubes can be cut into pieces of
desirable length \cite{el-hami03}, the techniques for unambiguous
determination of chirality of the walls have been successfully
demonstrated \cite{liu05}. All these gives us a cause for the
optimism that the proposed NEMS will be produced in the near
future.

\section{acknowledgement}
EB is indebted to the Royal Society for a UK Relocation
Fellowship. This work has been partially supported by the Russian
Foundation of Basic Research (AMP, YEL and ASK grants 05-02-17864
and 06-02-81036-Bel).


\begin{thebibliography}{10}
\bibitem{fennimore03} A.M. Fennimore, T.D. Yuzvinsky, W.Q. Han, M.S. Fuhrer, J. Cumings, A. Zettl,
Nature \textbf{424}, 408 (2003).
\bibitem{bourlon04} B. Bourlon, D.C. Glatti, L. Forro, A. Bachfold, Nano Lett., \textbf{4}, 709 (2004).
\bibitem{cumings00} J. Cumings, A. Zettl, Science, \textbf{289}, 602 (2000).
\bibitem{forro00} L. Forro, Science, \textbf{289}, 5479 (2000).
\bibitem{zheng02} Q. Zheng, Q. Jiang, Phys. Rev. Lett., \textbf{88}, 045503 (2002).
\bibitem{dequesnes02} M. Dequesnes, S.V. Rotkin, N.R. Aluru, Nanotechnology, \textbf{13}, 120–131 (2002).
\bibitem{kinaret03} J.M. Kinaret, T. Nord, S. Viefers, Appl. Phys. Lett., \textbf{82}, 1287 (2003).
\bibitem{maslov06} L. Maslov, Nanotechnology, \textbf{17}, 2475 (2006).
\bibitem{lozovik03} Yu.E. Lozovik, A.V. Minogin and A.M. Popov, Phys. Lett., \textbf{313}, 112 (2003).
\bibitem{lozovik03a} Yu.E. Lozovik, A.V. Minogin, A.M. Popov, JETP Letters, \textbf{77} 631
(2003).
\bibitem{bichoutskaia06a} E. Bichoutskaia, A.M. Popov, M.I. Heggie, Yu.E. Lozovik, Fullerenes,
 Nanotubes and Carbon Nanostructures, \textbf{14}, 131 (2006).
\bibitem{belikov04} A.V. Belikov, Yu.E. Lozovik, A.G. Nikolaev, A.M. Popov,
 Chem. Phys. Lett., \textbf{385},  72 (2004).
\bibitem{bichoutskaia06} E. Bichoutskaia, A.M. Popov, M.I. Heggie,  Yu.E. Lozovik,
Phys. Rev. B, \textbf{73}, 045435 (2006).
\bibitem{briddon00} P.R. Briddon, R. Jones, Phys. Stat. Sol., \textbf{217},  131  (2000).
\bibitem{bichoutskaia05} E. Bichoutskaia, A.M. Popov, A. El-Barbary,
M.I. Heggie, Yu.E. Lozovik, Phys. Rev. B, \textbf{71}, 113403
(2005).
\bibitem{perdew} J. P. Perdew, Y. Wang, Phys. Rev. B, \textbf{45}, 13244 (1992).
\bibitem{q-chem} J. Kong, C. A. White, A. I. Krylov, et al, J. Comput. Chem., \textbf{21} 1532 (2000).
\bibitem{girifalco00} L.A. Girifalco, M. Hodak, R.S. Lee,
Phys. Rev., \textbf{62}, 13104 (2000).
\bibitem{ellis00} D.E. Ellis, K.C. Mundimb, D. Fuksb, et al, Mater. Sc. in Semicond. Processing, \textbf{3}, 123 (2000).
\bibitem{degroot} M.H.DeGroot, M.J. Schervish, Probability and Statistics (3rd Edition), Addison-Wesley 2002, 816p.
\bibitem{grace04} I.M. Grace, S.W. Bailey, C.J. Lambert, Phys. Rev. B, \textbf{70}, 153405 (2004).
\bibitem{gao05} B. Gao, Y.F. Chen, M.S. Fuhrer, D.C. Glattli, A. Bachtold, Phys. Rev.
Lett. \textbf{95} 196802 (2005).
\bibitem{lebedev06} G.S. Ivanchenko, N.G. Lebedev, Phys. Solid State, \textbf{70}(1), in print (2007).
\bibitem{therm} E. Bichoutskaia, A.M. Popov,Yu.E. Lozovik, G.S. Ivanchenko, N.G.Lebedev, Phys. Lett. A, in print (2007).
\bibitem{bichoutskaia06b} E. Bichoutskaia, A.M. Popov, M.I. Heggie, Yu.E. Lozovik, Fullerenes,
 Nanotubes and Carbon Nanostructures, \textbf{14}, 215 (2006).
\bibitem{lozovik00} Yu.E. Lozovik, A.M. Popov, Chem. Phys. Lett., \textbf{328}, 355 (2000).
\bibitem{lozovik05} Yu.E. Lozovik, A.G. Nikolaev, A.M. Popov, Int. J. of Nanotechnology, \textbf{2}, 255 (2005).
\bibitem{lozovik06} Yu.E. Lozovik, A.G. Nikolaev, A.M. Popov, JETP, \textbf{103}, 447 (2006).
\bibitem{yu99} M.F. Yu, M.J. Dyer, G.D. Skidmore, H.W. Rohrs, X.K. Lu, K.D. Ausman, J.R. Von Ehr,
R.S. Ruoff, Nanotechnology \textbf{10}, 244 (1999).
\bibitem{tsang93} S.C. Tsang, P.J.F. Harris, M.L.H. Creen, Nature \textbf{362}, 520
(1993).
\bibitem{el-hami03} K. El-Hami, K. Mitsushige, Int. J. of Nanoscience \textbf{2}, 125 (2003).
\bibitem{liu05} Z. Liu, L.C. Qin, Chem. Phys. Lett. \textbf{408}, 75 (2005).
\end{thebibliography}
\end{document}